\documentclass[journal]{IEEEtran}

\usepackage[noadjust]{cite}

\usepackage{amssymb}
\usepackage[pdftex]{graphicx}
\graphicspath{{../pdf/}{../jpeg/}{./eps/}}
\DeclareGraphicsExtensions{.pdf,.jpeg,.png,.eps}

\usepackage{threeparttable}
\usepackage{amsmath}
\usepackage{array}
\usepackage{url}

\makeatletter
\def\endthebibliography{%
  \def\@noitemerr{\@latex@warning{Empty `thebibliography' environment}}%
  \endlist
}
\makeatother

\begin{document}
\title{Projection onto Epigraph Sets for Rapid Self-Tuning Compressed Sensing MRI}

\author{Mohammad~Shahdloo,
        Efe~Ilicak,
        Mohammad~Tofighi,
        Emine~U.~Saritas,~\IEEEmembership{Member,~IEEE,}
        A.~Enis~\c{C}etin,~\IEEEmembership{Fellow,~IEEE,}
        and~Tolga~\c{C}ukur*,~\IEEEmembership{Senior Member,~IEEE}
\thanks{This work was supported in part by TUBITAK 1001 Grant (117E171), by a
European Molecular Biology Organization Installation Grant (IG 3028),
by a TUBA GEBIP fellowship, and by BAGEP 2016 and BAGEP 2017 awards of
the Science Academy.
\textit{Asterisk indicates the corresponding author.}}
\thanks{M. Shahdloo, E. Ilicak, E. U. Saritas, and T. \c{C}ukur are with the Department of Electrical and Electronics Engineering and the National Magnetic Resonance Research Center, Bilkent University, Bilkent, 06800 Ankara, Turkey (e-mail: shahdloo@ee.bilkent.edu.tr; ilicak@ee.bilkent.edu.tr; saritas@ee.bilkent.edu.tr; cukur@ee.bilkent.edu.tr).}
\thanks{A. E. \c{C}etin is with the Department of Electrical and Computer Engineering, University of Illinois at Chicago, 1020 Sciences and Engineering Offices (SEO), 851 South Morgan St.(M/C 154) Chicago, IL 60607, on leave from the Department of Electrical and Electronics Engineering, Bilkent University, Ankara, Turkey (e-mail:  cetin@bilkent.edu.tr)}%
\thanks{M. Tofighi is with the Department of Electrical Engineering, Pennsylvania State University, 104 EE East Building, State College, PA (email: tofighi@psu.edu)}
\thanks{Copyright \copyright~2018 IEEE. Personal use of this material is permitted. However, permission to use this material for any other purposes must be obtained from the IEEE by sending a request to pubs-permissions@ieee.org.}}
%
%

\markboth{Citation information: DOI 10.1109/TMI.2018.2885599, IEEE Transactions on Medical Imaging}%
{Shahdloo \MakeLowercase{\textit{et al.}}: Projection onto Epigraph Sets for Rapid Self-Tuning Compressed Sensing MRI}
%

\IEEEpubid{\begin{minipage}{\textwidth}\ \\[12pt] \centering
  0278-0062 \copyright 2018 IEEE. Personal use is permitted, but republication/redistribution requires IEEE permission.\\
  See http://www.ieee.org/publications standards/publications/rights/index.html for more information.
\end{minipage}}


\maketitle
\vspace{-8pt}
\begin{abstract}
The compressed sensing (CS) framework leverages the sparsity of MR images to reconstruct from undersampled acquisitions. 
CS reconstructions involve one or more regularization parameters that weigh sparsity in transform domains against fidelity to acquired data. 
While parameter selection is critical for reconstruction quality, the optimal parameters are subject and dataset specific. 
Thus, commonly practiced heuristic parameter selection generalizes poorly to independent datasets. 
Recent studies have proposed to tune parameters by estimating the risk of removing significant image coefficients. 
Line searches are performed across the parameter space to identify the parameter value that minimizes this risk. 
Although effective, these line searches yield prolonged reconstruction times.  
 Here, we propose a new self-tuning CS method that uses computationally efficient projections onto epigraph sets of the $\ell_1$ and total-variation norms to simultaneously achieve parameter selection and regularization. 
 In vivo demonstrations are provided for balanced steady-state free precession, time-of-flight, and T1-weighted imaging. 
 The proposed method achieves an order of magnitude improvement in computational efficiency over line-search methods while maintaining near-optimal parameter selection.
\end{abstract}


%
\IEEEpeerreviewmaketitle

\section{Introduction}

\IEEEPARstart{T}{he} compressed sensing (CS) framework was recently proposed for accelerated MRI, where compressibility of MR images are employed to reconstruct from undersampled acquisitions \cite{Lustig_Sparse_2007,Block_Undersampled_2007,Liu_SparseSENSE_2008,ukur_Accelerated_2015}. 
To do this, CS reconstructions are typically cast as regularized optimization problems that weigh data consistency against sparsity in some transform domain (e.g., wavelet domain, total variation (TV)) \cite{Lustig_Sparse_2007}. 
The weighing between data consistency and sparsity is governed by regularization parameters.
High parameter values overemphasize sparsity at the expense of introducing inconsistency to acquired data samples, potentially leading to feature losses. 
Meanwhile, low parameter values render the reconstructions ineffective in suppressing residual aliasing and noise in undersampled acquisitions. 
Since the optimal regularization parameters are subject and dataset specific, time-consuming and potentially erroneous heuristic selection is performed in many studies, limiting the clinical utility of CS-MRI.

\begin{figure*}[!t]
\centering
\includegraphics{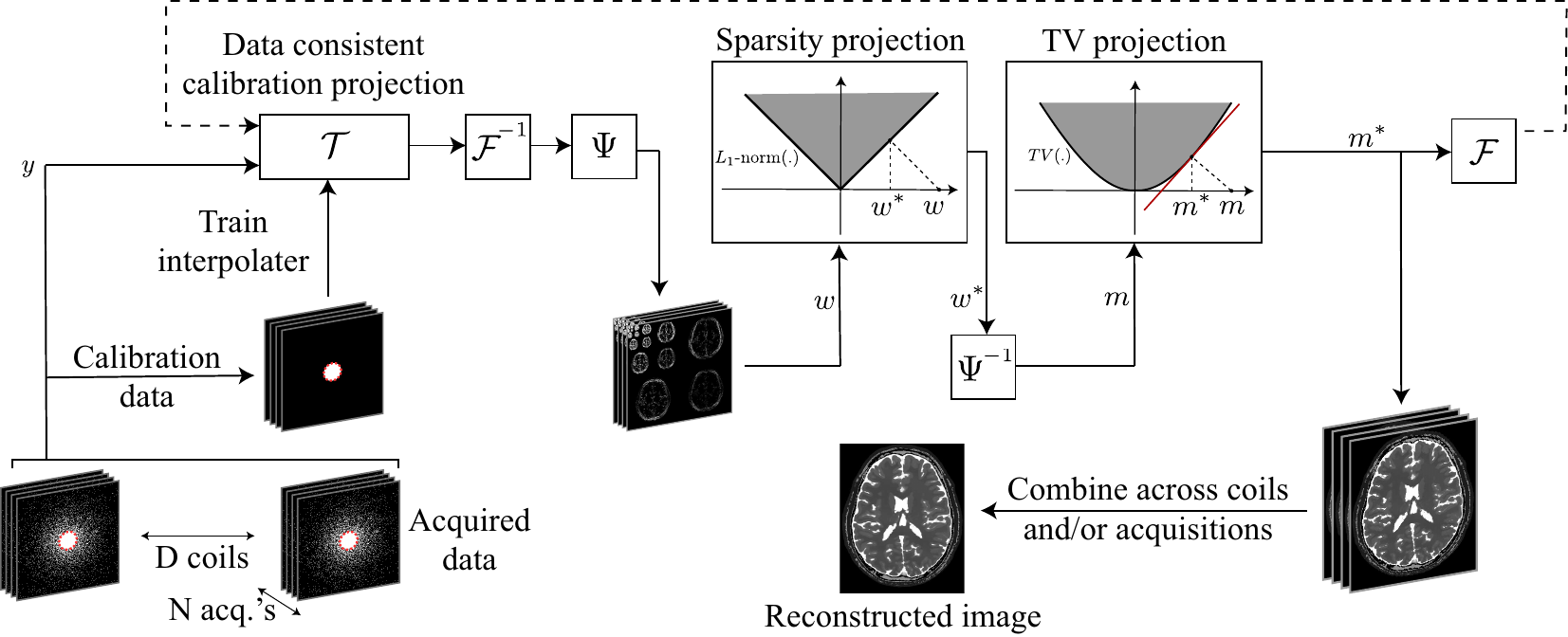}
\setlength{\abovecaptionskip}{-0cm}
\caption{Flowchart of the PESCaT reconstruction. 
PESCaT employs an alternating projections onto sets approach with three subprojections: data-consistent calibration projection, sparsity projection, and TV projection. 
The calibration projection linearly synthesizes unacquired k-space samples via a tensor interpolating kernel. 
The sparsity projection jointly projects wavelet coefficients of the multi-coil, multi-acquisition images onto the epigraph set of the $\ell_1$-norm function.
 The TV projection projects image coefficients onto the epigraph set of the TV-norm function. 
 These projections are performed iteratively until convergence. 
 Lastly, reconstructed images are combined across multiple coils and/or acquisitions. }
\label{fig:method}

\end{figure*}

Several unsupervised methods have been proposed to address parameter selection in CS-MRI. 
Empirical methods including the L-curve criterion (LCC) follow the notion that the optimal parameter should be selected to attain a favorable trade-off between data consistency and regularization objectives \cite{Hansen_Analysis_1992,Hansen_The_1993,Regiska_A_1996}. 
 Assuming this trade-off is characterized by an L-shaped curve, LCC selects the parameter on the point of maximum curvature \cite{Zhang_Adaptive_1998}. 
 LCC has been successfully demonstrated for parameter selection in several applications including parallel imaging \cite{Lin_Parallel_2004,Ying_On_2004}, quantitative susceptibility mapping \cite{Bilgic_Sparse_2015}, and diffusion spectrum imaging \cite{Bilgic_Fast_2014}.
 However, curvature assessment is computationally inefficient and typically sensitive to numerical perturbation and nonlinearities in the reconstruction problem \cite{Vogel_Non_1996,Kressler_Nonlinear_2010,Giryes_The_2011}.

Alternatively, parameters can be selected based on analytical estimates of the reconstruction error to optimize the regularization parameters.
These methods include generalized cross-validation (GCV) \cite{Golub_Generalized_1979}, and methods based on Stein's unbiased risk estimator (SURE) \cite{Stein_Estimation_1981,Zhang_Adaptive_1998}. 
In GCV, an analytical measure for reconstruction error is estimated that asymptotically converges to the true error \cite{Golub_Generalized_1979}.
The GCV measure is derived as a function of the sampling pattern, regularization function, and regularization parameter. 
Parameter estimation via minimization of the GCV measure has been used in a variety of applications such as functional MRI \cite{Carew_Optimal_2003}, perfusion imaging \cite{Sourbron_Choice_2004}, and dynamic MRI \cite{sourbron2004deconvolution}.
However, the GCV measure can be expensive to compute and yields biased estimates of the true error with limited number of data samples \cite{Ramani_Regularization_2012}.

\IEEEpubidadjcol

A recent approach instead uses the SURE criterion to estimate the expected value of the mean-square error (MSE) of the reconstruction.
Given a specific parameter value and an estimate of the noise variance, Stein's lemma \cite{Stein_Estimation_1981} is used to compute online estimates of MSE. 
Subsequently, a line search over potential parameter values is performed for selecting the optimal parameter at each iteration.
SURE-based parameter selection has produced promising results in several sparse recovery applications including CS-MRI \cite{Luisier_A_2007,Blu_The_2007,Ville_Nonlocal_2011,Khare_Accelerated_2012,Deledalle_Stein_2014,Guo_Near_2015}. 
Unfortunately, parameter searches that need to be performed in each iteration cause substantial computational burden. 

Here we introduce a computationally efficient self-tuning reconstruction method, named PESCaT (Projection onto Epigraph Sets for reconstruction by Calibration over Tensors), that can handle both single-acquisition and multi-acquisition datasets. 
To jointly reconstruct undersampled acquisitions, PESCaT performs tensor-based interpolation across acquired data, complemented by sparsity regularization of wavelet coefficients and TV regularization of image coefficients.

Since wavelet coefficients show varying sparsity across subbands and decomposition levels, PESCaT uses different $\ell_1$ regularization parameters for each subband and level. 
Similarly, multi-coil multi-acquisition image coefficients may show varying spatial gradients, so different TV regularization parameters are used for each coil and acquisition. 
Parameters are efficiently tuned via simple geometric projections onto the boundary of the convex epigraph sets for the $\ell_1$- and TV-norm functions. 
This formulation transforms the selection of many different regularization parameters for multiple subbands, levels, coils, and acquisitions into the selection of two scaling factors for the $\ell_1$-norm and TV-norm epigraphs. 
These factors can be reliably tuned on training data, yielding consistent performance across sequences, acceleration factors, and subjects. 
Comprehensive demonstrations on simulated brain phantoms, and in vivo balanced steady-state free-precession (bSSFP), T1-weighted, and angiographic acquisitions indicate that PESCaT enables nearly an order of magnitude improvement in computational efficiency compared to SURE-based methods, without compromising reconstruction quality.  

\section{Theory}
Our main aim is to introduce a fast joint reconstruction method that automatically selects the free parameters for regularization terms based on $\ell_1$- and TV-norms. 
We consider the application of this self-tuning reconstruction to single-coil multi-acquisition, multi-coil single-acquisition, and multi-coil multi-acquisition MRI datasets. 
In the following sections, we introduce the regularized reconstruction problem, and its solution via projection onto epigraph sets for unsupervised parameter selection. 
 \vspace{-8pt}
\subsection{Reconstruction by calibration over tensors}
Compressive sensing (CS) techniques proposed for static MRI acquisitions typically leverage encoding information provided either by multiple coils \cite{Lustig_SPIRiT_2010,Block_Undersampled_2007,Liang_Accelerating_2009} or by multiple acquisitions \cite{Ilicak_Profile_2017,bilgic2011multi,huang2014fast} to enable recovery of unacquired data samples. 
Yet, simultaneous use of information across coils and acquisitions can benefit phase-cycled bSSFP \cite{Biyik_Reconstruction_2017,hilbert2017true}, multi-contrast \cite{MRM:MRM25142,Toygan,Berkin_new} or parametric imaging \cite{velikina2013accelerating,zhao2015accelerated,zhang2015accelerating}. 
Here we consider a joint reconstruction framework for multi-coil, multi-acquisition datasets, based on a recent method that we have proposed named ReCaT (Reconstruction by Calibration over Tensors) \cite{Biyik_Reconstruction_2017}. 
ReCaT rests on the following spatial encoding model for the signal measured in acquisition $n\in [1,\dots,N]$ and coil $d\in [1,\dots,D]$:
\begin{align}
S_{nd}(r) = P_n(r)C_d(r)S_0(r),
\end{align}
where $r$ is the spatial location, $P_n$ is the acquisition spatial profile, $C_d$ is the coil sensitivity profile, and $S_0$ is the signal devoid of coil sensitivity and acquisition profile modulations. 
ReCaT seeks to linearly synthesize missing k-space samples from neighboring acquired samples across all coils and acquisitions. 
A tensor interpolation kernel is used for this purpose:
\begin{align}
x_{nd} = \sum_{i=1}^N\sum_{j=1}^Dt_{ij,nd}(k_r) \circledast x_{ij}(k_r),
\label{dd2}
\end{align}
where $x_{nd}$ is the k-space data from $n^{th}$ acquisition and $d^{th}$ coil, $k_r$ is the k-space location, and $\circledast$ is the convolution operation. 
Here $t_{ij,nd}(k_r)$ accounts for the contribution of samples from acquisition $i$ and coil $j$ to $x_{nd}$. 
Equation (\ref{dd2}) can be compactly expressed as:
\begin{align}
x = \mathcal{T}x.
\label{eq:interpolation}
\end{align}

\begin{figure}
\centering
 \includegraphics{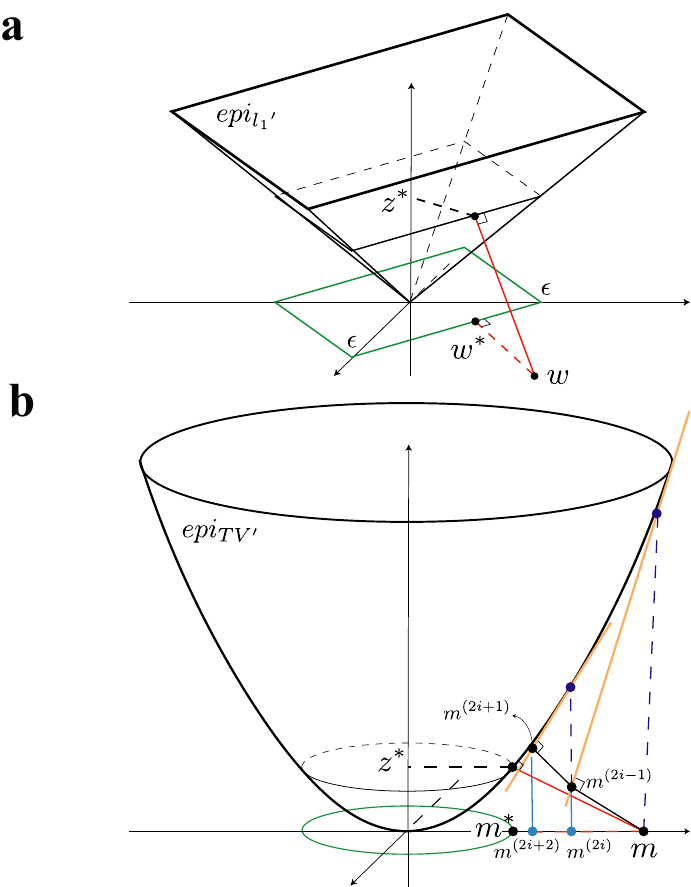}
\setlength{\abovecaptionskip}{-.1cm}
\caption{
The projection onto epigraph sets (PES) approach illustrated in $\mathbb{R}^3$. 
\textbf{(a)} PES for $\ell_1$-regularization. 
An input vector $w$ (e.g., vector of wavelet coefficients of image $m$) is projected onto the epigraph set of the $\ell_1$-norm function ($epi_{\ell_1'}$). 
This projection results in the output $[w^* z^*]^T$, thereby inherently calculating the projection of $w$ onto the $\ell_1$-ball in $\mathbb{R}^2$ ($w^*$).
 The size of the $\ell_1$-ball ($\epsilon$) depicted in green color depends on $z^*$.
\textbf{(b)} PES for TV regularization. 
Unlike PES-$\ell_1$, PES-TV has no closed-form solution, and is instead implemented via an iterative epigraphical splitting procedure. 
At the $i^{th}$ iteration, the input vector $m^{(2i-1)}$ is projected onto the supporting hyperplane (orange line), resulting in $m^{(2i+1)}$. 
This intermediate vector is then projected on the level set to compute $m^{(2i+2)}$. 
Through successive iterations the output gradually converges to the desired projection point on the epigraph set $[m^* z^*]^T$, thereby inherently calculating the projection of $m$ onto the TV-ball in $\mathbb{R}^2$ ($m^*$). 
The size of the TV-ball depends on $z^*$.}
\label{fig:projection}
\end{figure}

\begin{figure}[!t]
\centering
\includegraphics{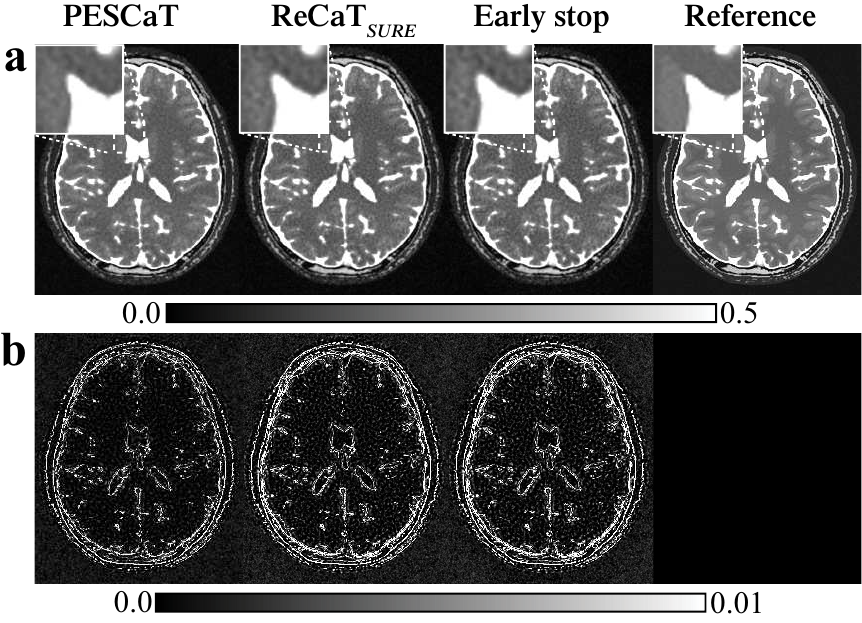}
\setlength{\abovecaptionskip}{-.3cm}
\caption{Reconstructions of phase-cycled bSSFP acquisitions of the simulated brain phantom.
Simulations assumed $\alpha$=$45^{\circ}$, TR/TE=5/2.5 ms, a fieldmap of $0\pm62$ Hz (mean$\pm$std), N = 6 phase-cycles and R = 6. 
\textbf{(a)} PESCaT, ReCaT\textsubscript{\textit{SURE}}, ReCaT\textsubscript{\textit{SURE}} with early stop, and the fully-sampled reference images are shown. 
White boxes display zoomed-in portions of the images. 
\textbf{(b)} Mean-squared error between the reconstructed and reference images are shown. 
PESCaT yields visibly reduced errors compared to both ReCaT\textsubscript{\textit{SURE}} and ReCaT\textsubscript{\textit{SURE}} with early stop.}
\label{fig:phantom}
\end{figure}

\begin{figure*}[!t]
\centering
\includegraphics{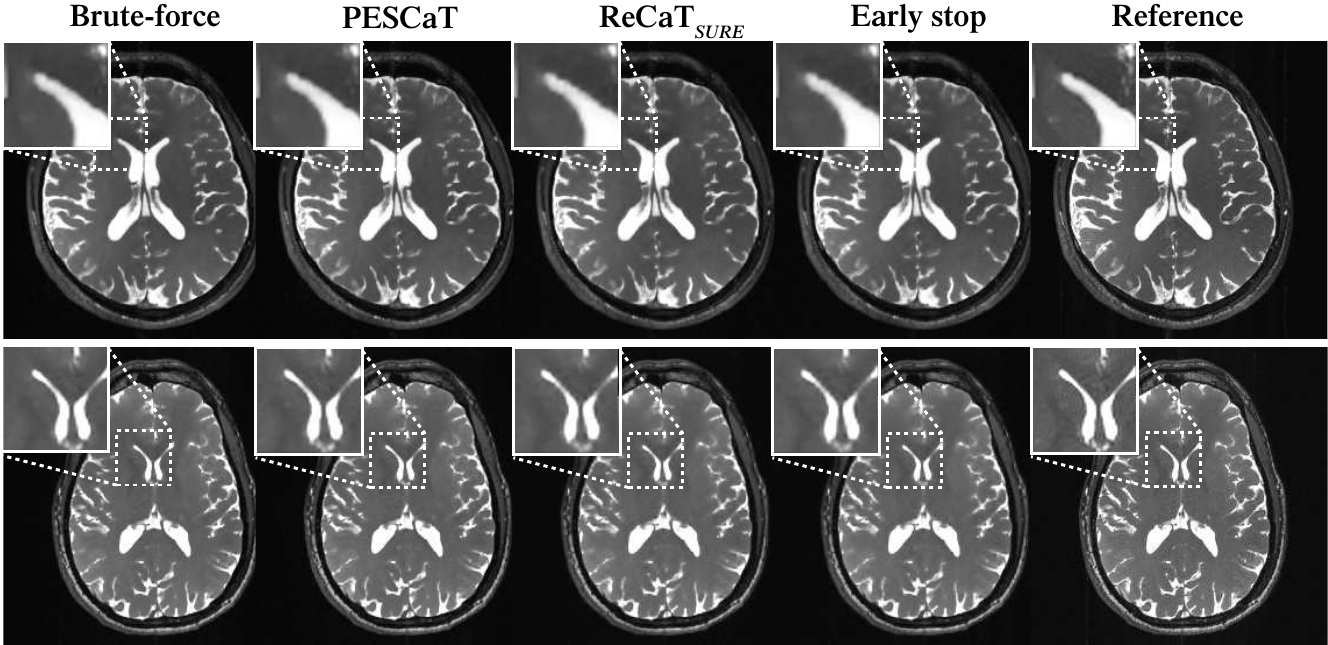}

\setlength{\abovecaptionskip}{-.1cm}
\caption{
Reconstructions of in vivo bSSFP acquisitions of the brain at R=6. 
Brute-force, PESCaT, ReCaT\textsubscript{\textit{SURE}}, ReCaT\textsubscript{\textit{SURE}} with early stop, and reference images are shown in two representative subjects. 
White boxes display zoomed-in portions of the images. 
PESCaT achieves significantly improved image quality compared to ReCaT\textsubscript{\textit{SURE}} with early stop that was matched PESCaT in terms of the total reconstruction time. 
Furthermore, PESCaT yields similar image quality to ReCaT\textsubscript{\textit{SURE}} and brute-force methods, while also maintaining greater computational efficiency.}
\label{fig:ssfp}
\end{figure*}
 \vspace{-8pt}
\subsection{PESCaT}
In the previous study where we proposed ReCaT, a basic implementation was considered that did not include any regularization terms to enforce sparsity \cite{Biyik_Reconstruction_2017}. 
Here we introduce an improved version, PESCaT, that incorporates sparsity and TV penalties:
\begin{align}
& \underset{{x}_{nd}}{\text{min}}  \Big\{ \sum_{n=1}^N\sum_{d=1}^D\left|\left|(\mathcal{T-I})x_{nd}\right|\right|_2^2 &\nonumber \\
&+ \sum_{l=1}^L \sum_{s=1}^3\lambda_{\ell_1,ls}\sum_{n=1}^N\sum_{d=1}^D \big|\big|\Psi_{ls}\big\{\mathcal{F}^{-1}\{x_{nd}\}\big\}\big|\big|_1  &\nonumber\\
&+ \sum_{n=1}^N\sum_{d=1}^D \lambda_{TV,nd}||\mathcal{F}^{-1}\{x_{nd}\}||_{TV}\Big\}, &
\label{pescat}
\end{align}
where $\Psi_{ls}$ is the wavelet operator for subband $s$ and level $l$, $\mathcal{I}$ is the identity operator, and $\mathcal{F}^{-1}$ is the inverse Fourier operator. 
A separate $\ell_1$-regularization parameter, $\lambda_{\ell_1,ls}$, is prescribed for each subband and level of the wavelet coefficients. 
Sparsity regularization is performed on the three high-pass subbands while the low-pass subband is kept intact to avoid over-smoothing. 
Meanwhile, a separate TV regularization parameter, $\lambda_{TV,nd}$, is used for each acquisition and coil.
Because wavelet coefficients are aggregated across the coil and acquisition dimensions, $\lambda_{\ell_1,ls}$ varies across wavelet levels and subbands but it is fixed across coils or acquisitions.

In this study, we implemented PESCaT in a constrained optimization formulation equivalent to the Lagrangian formulation in (\ref{pescat}): 
\begin{equation}
\begin{aligned}
& \underset{{x}_{nd}}{\text{min}}
& & \sum_{n=1}^N\sum_{d=1}^D\left|\left|(\mathcal{T-I})x_{nd}\right|\right|_2^2 \\
& \text{subject to}
& & \sum_{n=1}^N\sum_{d=1}^D \big|\big|\Psi_{ls}\big\{\mathcal{F}^{-1}\{x_{nd}\}\big\}\big|\big|_1  \leq \epsilon_{\ell_1,ls}\, \;\\
& & & s=1,2,3\\
& & & l=1,\dots,L; \\
& & & ||\mathcal{F}^{-1}\{x_{nd}\}||_{TV} \leq \epsilon_{TV,nd}\\
& & & n=1,\dots,N\\
& & & d=1,\dots,D \\
\end{aligned}
\label{pescatconst}
\end{equation}
where $\epsilon_{\ell_1,ls}$ are the constraints on the sparsity of the reconstruction, and $\epsilon_{TV,nd}$ are the constraints on the TV of the reconstruction. 
The optimization problem in (\ref{pescatconst}) was solved via an alternating projections onto sets algorithm. 
As outlined in Fig. \ref{fig:method}, this algorithm involves three consecutive projections, namely data-consistent calibration, sparsity, and TV projections. 
The calibration projection linearly synthesized unacquired k-space samples via the tensor interpolating kernel.
To perform this projection while enforcing strict consistency to acquired data, an iterative least-squares algorithm was employed \cite{Lustig_SPIRiT_2010}. 
The sparsity projection jointly projected wavelet coefficients of images onto the epigraph set of the $\ell_1$-norm function. 
The TV projection projected image coefficients onto the epigraph set of the TV-norm function. 
These projections were performed iteratively until convergence.
At each iteration, MSE between the reconstructed image in the current iterate and the previous iterate was first measured, and the percentage change in MSE across consecutive iterations was then calculated. 
Convergence was taken to be the iteration at which the percentage change in MSE fell below 20\%. 
Lastly, reconstructed images were combined across multiple coils and/or acquisitions.
Note that because PESCaT is structured modularly regarding the calibration, sparsity, and TV projections, it is trivial to implement variants that only employ sparsity or TV regularization. 

 \vspace{-8pt}
\subsection{Parameter tuning by projection onto epigraph sets}
Careful tuning of constraint parameters in (\ref{pescatconst}) is critical for a successful reconstruction. 
Selecting too tight constraints can lead to loss of important image features, whereas selecting too loose constraints will yield substantial residual noise and aliasing. 
When only a few parameters are to be tuned, an exhaustive search over a relevant range of values followed by visual inspection is typically exercised. 
However, even in a modest dataset with $D = 4$ coils and $N = 4$ acquisitions, and assuming $L = 4$ wavelet decomposition levels there are 28 distinct parameters involved in (\ref{pescatconst}). 
Thus, the exhaustive search approach is impractical. 

\begin{figure*}[!t]
\centering
\includegraphics{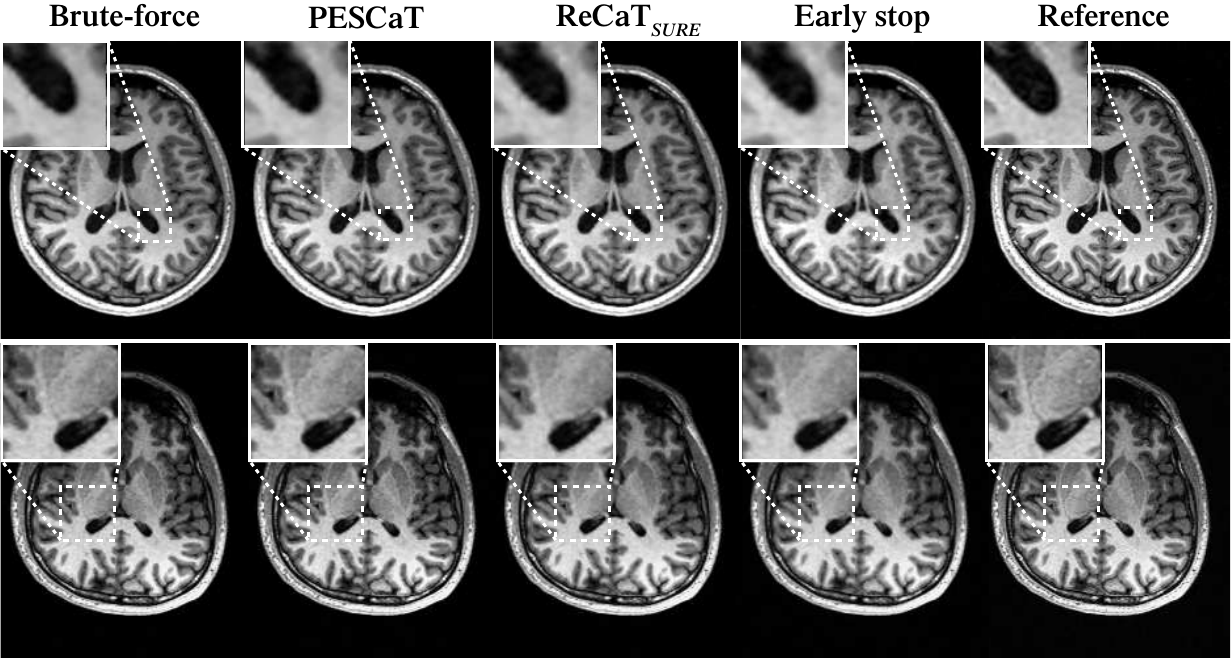}
\setlength{\abovecaptionskip}{-.1cm}
\caption{
Reconstructions of in vivo T1-weighted acquisitions of the brain at R=4. 
Brute-force, PESCaT, ReCaT\textsubscript{\textit{SURE}}, ReCaT\textsubscript{\textit{SURE}} with early stop, and reference images are shown in two representative subjects. 
White boxes display zoomed-in portions of the images. 
PESCaT achieves significantly improved image quality compared to ReCaT\textsubscript{\textit{SURE}} with early stop that was matched PESCaT in terms of the total reconstruction time. 
Meanwhile, PESCaT yields similar image quality to ReCaT\textsubscript{\textit{SURE}} and brute-force methods.}
\label{fig:t1}
\end{figure*}

In this study, we perform self-tuning of the constraint parameters in (\ref{pescatconst}) via projections onto epigraph sets of the respective regularization terms.
 Let $\mathcal{U} \in \mathbb{R}^k$ be a closed convex set, $\Phi:\mathbb{R}^{k}\rightarrow\mathbb{R}$ be a convex function (e.g., $\ell_1$-norm and TV-norm functions), and  $\hat{u}\in \mathbb{R}^k$ be an input vector (e.g., wavelet coefficients for $\ell_1$-norm or image coefficients for TV-norm). 
 The proximal operator of $\Phi^2$ is:
 \begin{align}
prox_{\Phi^2}(\hat{u})  = \arg\min_{u\in \mathcal{U}} ||\hat{u}-u||_2^2 +  \Phi^2(u),
\label{eq:proximal}
\end{align}
where $u$ is the auxiliary variable. 
We prefer to use $\Phi^2$ here since it allows us to express the solution as a simple geometric projection. Specifically, the problem in (\ref{eq:proximal}) can be stated in vector form by mapping onto $\mathbb{R}^{k+1}$:
\begin{align}
\min_{u\in\mathcal{U}}\left|\left| \begin{bmatrix} \hat{u} \\ 0  \end{bmatrix} - \begin{bmatrix} u \\ \Phi(u) \end{bmatrix}\right|\right|_2^2.
\label{eq:vecproximal}
\end{align}
Here we propose to implement the proximal operator in (\ref{eq:proximal}) by identifying the closest vector $\begin{bmatrix} u^* &\Phi(u^*)  \end{bmatrix}^T \in \mathbb{R}^{k+1}$ to $\begin{bmatrix} \hat{u} &0  \end{bmatrix}^T$.
 This solution can be shown to be equivalent to the orthogonal projection of the vector $\begin{bmatrix} \hat{u} &0  \end{bmatrix}^T$ onto the epigraph set of $\Phi$ ($epi_{\Phi}$) defined as:
 \begin{align} 
epi_{\Phi} = \{ \begin{bmatrix} u & z \end{bmatrix}^T: z \ge\Phi(u)\},
\end{align}
where $z$ denotes an upper bound for the function $\Phi(u)$.
The projection onto $epi_{\Phi}$ is the closest solution to $\hat{u}$ that lies on the boundary of the epigraph set. 
Since the epigraph set of a convex function is also convex, this projection will yield the global optimum solution. 
Note that projections onto the epigraph set will yield the solution of the proximal operator only if the search space of the proximal operator is a convex set $\mathcal{U} \in \mathbb{R}^k$  \cite{Cetin:wh}.
In practice, a family of solutions can be obtained by introducing a scaling parameter to alter the size of the epigraph set:
\begin{align} 
epi_{\Phi'} = \{\begin{bmatrix} u & z \end{bmatrix}^T: z \ge\beta_{\Phi}\Phi(u)\}.
\label{eq:scaledepi}
\end{align}
Here, $\beta_{\Phi}$ serves to control the allowed degree of deviation of $u^*$ from $\hat{u}$. Note that both $z^*$ and $u^*$ are computed via an orthogonal projection of the input onto $epi_{\Phi}'$. Since the scales of $z^*$ and $u^*$ vary proportionately to the scale of $\hat{u}$, $\beta_{\Phi}$ can be described in absolute terms. $\beta_{\Phi} > 1$ scales down the epigraph set, resulting in a solution $u^*$ that deviates further from $\hat{u}$. 
Meanwhile, $0<\beta_{\Phi}<1$ scales up $epi_{\Phi}$, resulting in a solution $u^*$ that is closer to $\hat{u}$, where $u^*=\hat{u}$ as $\beta_{\Phi}\rightarrow0$.
To obtain more conservative solutions, here we used $0<\beta_{\Phi}<1$ for both sparsity and TV projections. 
The resulting projection point determines both the size of the $\Phi$-ball in $\mathbb{R}^{k}$ (i.e. $\ell_1$-ball or TV-ball, see Fig. 2) and the actual projection onto the ball.
Hence, the proximal operator in (\ref{eq:proximal}) enables assessing the optimal constraint parameters in (\ref{pescatconst}) using the input vector $\hat{u}$ as explained below.

 \vspace{-0pt}
 \subsubsection{Self-tuning sparsity projection}
The sparsity projections were implemented using projections onto the epigraph set of the $\ell_1$-norm function, applied on wavelet coefficients. 
The image coefficients $m_{nd} = \mathcal{F}^{-1}\{x_{nd}\}$ are obtained by inverse Fourier transformation of k-space data, $x_{nd}$, for acquisition $n$ and coil $d$. 
The wavelet coefficients for $m_{nd}$ are then given by $w_{ls,nd} = \Psi_{ls}\{m_{nd}\}$ at subband $s$ and level $l$, and $w_{ls}$ denotes the aggregate vector pooling $w_{ls,nd}$ across coils and acquisitions. 
Assuming $\hat{w} = w_{ls}$ is the input vector, the proximal formulation in (\ref{eq:proximal}) becomes:
\begin{align}
prox_{\ell_1^2}(\hat{w})  = \arg\min_{u} ||\hat{w}-u||_2^2 + ||u||_1^2.
\label{eq:proximall1}
\end{align}
The solution to (\ref{eq:proximall1}) is then obtained by projecting $\begin{bmatrix} w_{ls}&0  \end{bmatrix}^T$ onto the scaled epigraph set (see Fig. \ref{fig:projection}a):
\begin{align}
 epi_{\ell_1'} = \{ \begin{bmatrix} u &  z \end{bmatrix}^T \in \mathbb{R}^{k+1} :  z \ge \beta_{\ell_1}||u||_1  \},
 \end{align}
 where $\beta_{\ell_1}$ denotes the epigraph scaling factor for the $\ell_1$-norm.
As demonstrated in Fig. \ref{fig:projection}a, the closest orthogonal projection of $\begin{bmatrix} w_{ls}&0  \end{bmatrix}^T$ onto the epigraph set lies on the boundary of $epi_{\ell_1'}$. 
For the simple case of $\mathbb{R}^2$ ($k=1$), $\begin{bmatrix} w_{ls}&0  \end{bmatrix}^T$ is projected onto the $z = \beta_{\ell_1}|u|$ line, yielding $z^*_{ls} = \frac{\beta_{\ell_1}|w_{ls}|}{\beta_{\ell_1}^2+1}$.
 It can be shown that for arbitrary $k$, the $z$-intercept is:
\begin{align}
z^*_{ls} = \dfrac{\beta_{\ell_1}||w_{ls}||_1}{\beta_{\ell_1}^2k+1}.
\end{align}
The value of the z-intercept also determines the size of the respective $\ell_1$-ball, $ B_{\ell_1,ls} = \{u\in \mathbb{R}^k : ||u||_1 \le  \epsilon_{\ell_1,ls} \}$, as:
 \begin{align}
 \epsilon_{\ell_1,ls}  = \dfrac{z^*_{ls}}{\beta_{\ell_1}}.
\end{align}\begin{figure*}[!t]
\centering
\includegraphics{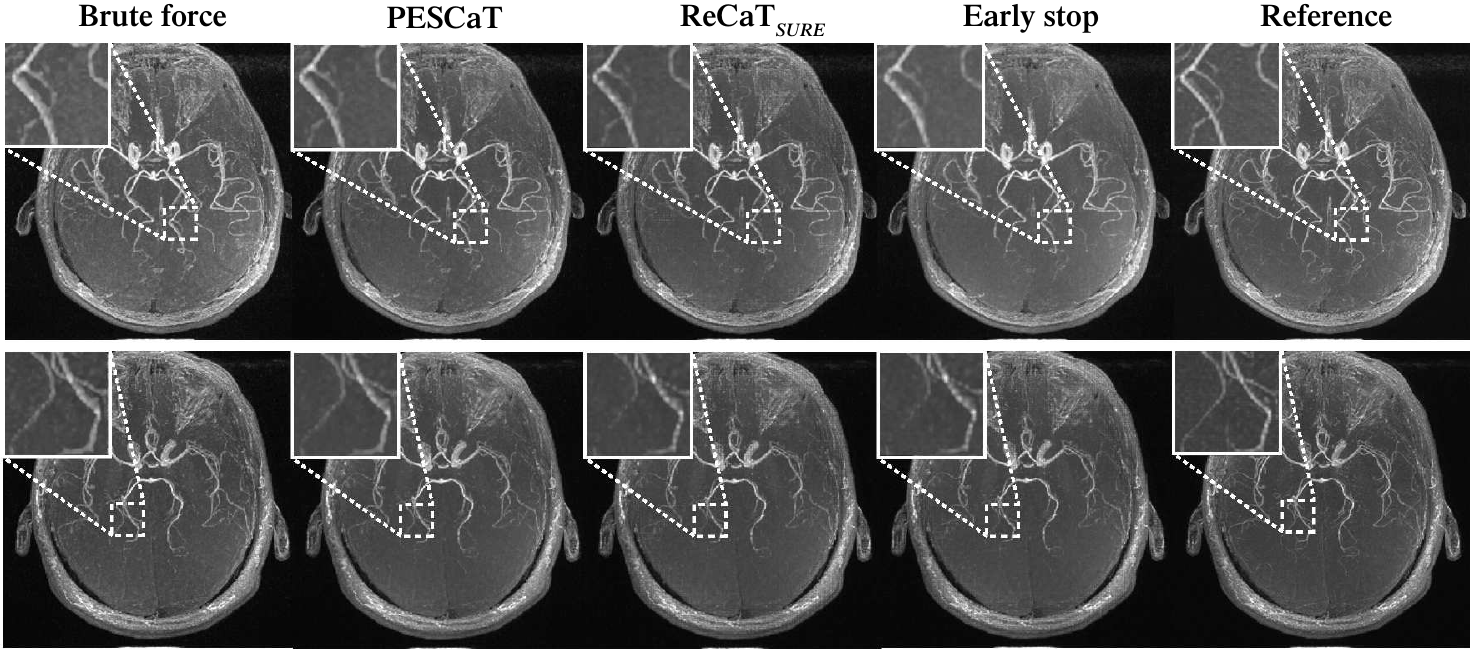}
\setlength{\abovecaptionskip}{-.1cm}
\caption{
Reconstructions of in vivo ToF angiography acquisitions of the brain at R=4. 
Maximum-intensity projection (MIP) views of brute-force, PESCaT, ReCaT\textsubscript{\textit{SURE}}, ReCaT\textsubscript{\textit{SURE}} with early stop, and reference brain volumes are shown in two representative subjects. 
White boxes display zoomed-in portions of the MIPs. 
PESCaT yields superior depiction of vasculature compared to both ReCaT\textsubscript{\textit{SURE}} and ReCaT\textsubscript{\textit{SURE}} with early stop.
 It also maintains similar image quality to brute-force reconstructions.}
\label{fig:tof}
\end{figure*}Therefore, $w^*_{ls}$ can be computed by finding the projection of $w_{ls}$ onto the $\ell_1$-ball of size $\epsilon_{\ell_1,ls}$.
 To efficiently implement this projection, we used a soft-thresholding operation \cite{khare2010multilevel}:
\begin{align}
w^*_{ls} = e^{i\angle w_{ls}}\max(|w_{ls}|-\theta_{ls},0),
\label{eq:soft-thresh}
\end{align}
where magnitudes of wavelet coefficients are subjected to a threshold of $\theta_{ls}$, and phases of coefficients are individually restored via $e^{i\angle w_{ls}}$.
We propose to determine the value of $\theta_{ls}$ given $\epsilon_{\ell_1,ls}$ using an efficient ranking algorithm \cite{duchi2008efficient}. 
The proposed algorithm first sorts the absolute values of the wavelet coefficients $w_{ls,nd}$ to attain a rank-ordered sequence $\{\mu_i\}_{i=1}^k$ where $\mu_1>\mu_2>\ldots>\mu_k$. 
This sequence is then analyzed to find the threshold that approximately yields a resultant $\ell_1$-norm of value $\epsilon_{\ell_1,ls}$ in the thresholded coefficients:
\begin{flalign}
&\rho_{ls}=\max\{j\in \{1,2,\ldots,k\}: \mu_j-\dfrac{1}{j}(\sum_{r=1}^j\mu_r -\epsilon_{\ell_1,ls})>0\},&\nonumber\\
&\theta_{ls} = \dfrac{1}{\rho_{ls}}(\sum_{n=1}^{\rho_{ls}}\mu_n -\epsilon_{\ell_1,ls}).&
\end{flalign}
Note that the determined threshold directly translates to $\lambda_{\ell_1,ls}$ in (\ref{pescat}) by \cite{Parikh:2014bs}:
\begin{align}
\lambda_{\ell_1,ls} =  2\theta_{ls}.
\end{align}
Projections were separately performed for each subband $s$ at each wavelet decomposition level $l$ to determine the respective $w^*_{ls}$, and $\epsilon_{\ell_1,ls}$.
 Since wavelet coefficients were pooled across coils and acquisitions, parameter selection is performed jointly across coils and acquisitions.
 Since the only free parameter in the proposed method is the epigraph scaling constant $\beta_{\ell_1}$, the selection of $3\times L$ parameters in (\ref{pescatconst}) are transformed into the selection of a single parameter. 
 Here, the optimal $\beta_{\ell_1}$ was empirically determined in a group of training subjects and then used to obtain reconstructions in held-out test subjects. 
 \vspace{-0pt}
 \subsubsection{Self-tuning TV projection}
 The TV projections were implemented using projections onto the epigraph set of the TV-norm function, applied on image coefficients. 
Letting $\hat{m} = m_{nd}$ be the input vector, the proximal formulation in (\ref{eq:proximal}) becomes:
\begin{align}
prox_{TV^2}(\hat{m})  = \arg\min_{u} ||\hat{m}-u||_2^2 + ||u||_{TV}^2.
\label{eq:proximaltv}
\end{align}
The solution to (\ref{eq:proximaltv}) is then obtained by projecting $\begin{bmatrix} m_{nd}&0  \end{bmatrix}^T$ onto the scaled epigraph set (see Fig. \ref{fig:projection}b):
\begin{align}
 epi_{TV'} = \{ \begin{bmatrix} u &  z \end{bmatrix}^T \in \mathbb{R}^{k+1} :  z \ge \beta_{TV}||u||_{TV}  \},
 \end{align}
 where $\beta_{TV}$ denotes the epigraph scaling factor for the TV-norm.
Unlike the projection onto the $\ell_1$-norm epigraph, projection onto generic epigraph sets (including TV-norm epigraph) does not have a closed-form solution. 
As demonstrated in Fig. \ref{fig:projection}b, PESCaT uses an iterative epigraphical splitting method to perform the projection efficiently \cite{Tofighi_Signal_2014}.
In the initial step of this approach, complex-valued $m^{(0)}=m_{nd}$ is projected onto the supporting hyperplane of $epi_{TV'}$ at $\begin{bmatrix} m_{nd} &  \beta_{TV}||m_{nd}||_{TV} \end{bmatrix}^T$ resulting in $m^{(1)}$.
The supporting hyperplane is determined by evaluating the gradient of the epigraph surface.
In the following step, $m^{(1)}$ is projected onto the level set, $
L_{TV} = \{  \begin{bmatrix} u &  z \end{bmatrix}^T :  z \le 0\}
$, by forcing the last element of $m^{(1)}$ to zero.
This projection yields the next estimate $m^{(2)}$.
These two projections are iterated.
Note that all steps of the splitting procedure are performed in complex domain, thereby, regularizing magnitude and phase channels simultaneously.
Previous studies have shown that the second derivative of distance between the input vector and the projections on the supporting hyperplanes ($||m_{nd} - m^{(2i+1)}||_2$) is negative as the projections approach to the true projection solution and is positive as the projections deviate from it \cite{Cetin:wh}. 
Thus, in case of a sign change in the second derivative a refinement step is performed, where $m^{(2i)}$ is projected onto the supporting hyperplane at $\frac{m^{(2i+1)}+m^{(2i-1)}}{2}$.
This heuristic approach has been shown to converge to the global solution for TV projections \cite{Tofighi_Signal_2014}.
Note that the projection uniquely specifies the z-intercept, $z^*_{nd}$.
Hence, the size of the corresponding TV-ball, 
$
 B_{TV,nd} = \{u\in \mathbb{R}^k : ||u||_{TV} \le \epsilon_{TV,nd}\}, 
$
can be calculated as:
\begin{align}
\epsilon_{TV,nd} = \dfrac{z^*_{nd}}{\beta_{TV}}.
\end{align}
Note that it is nontrivial to explicitly express $\lambda_{TV,nd}$ in (\ref{pescat}) in terms of $\epsilon_{TV,nd}$ in (\ref{pescatconst}). 
Yet, constraining $\epsilon_{TV,nd}$ implicitly enforces a set of regularization parameters $\lambda_{TV,nd}$.

Projections were separately performed for each acquisition $n$ and coil $d$ to determine the respective $m^*_{nd}$, and $\epsilon_{TV,nd}$.
 Since the only free parameter is the epigraph scaling constant $\beta_{TV}$, the selection of $N\times D$ parameters in (\ref{pescatconst}) is transformed into the selection of a single parameter. 
 Here, the optimal $\beta_{TV}$ was empirically determined in a group of training subjects and then used to obtain reconstructions in held-out test subjects. 

 All reconstruction algorithms were executed in MATLAB (MathWorks, MA). 
 The implementations used libraries from the {SPIRiT} toolbox \cite{Lustig_SPIRiT_2010}. 
 The PESCaT algorithm is available for general use at \url{http://github.com/icon-lab/mrirecon}.
\vspace{-4pt}
\section{Methods}
 \vspace{-3pt}
\subsection{Alternative reconstructions}
To demonstrate the performance of PESCaT, we compared it against several alternative reconstructions that aim to select regularization parameters. 

\begin{figure*}[!t]
\centering
\includegraphics{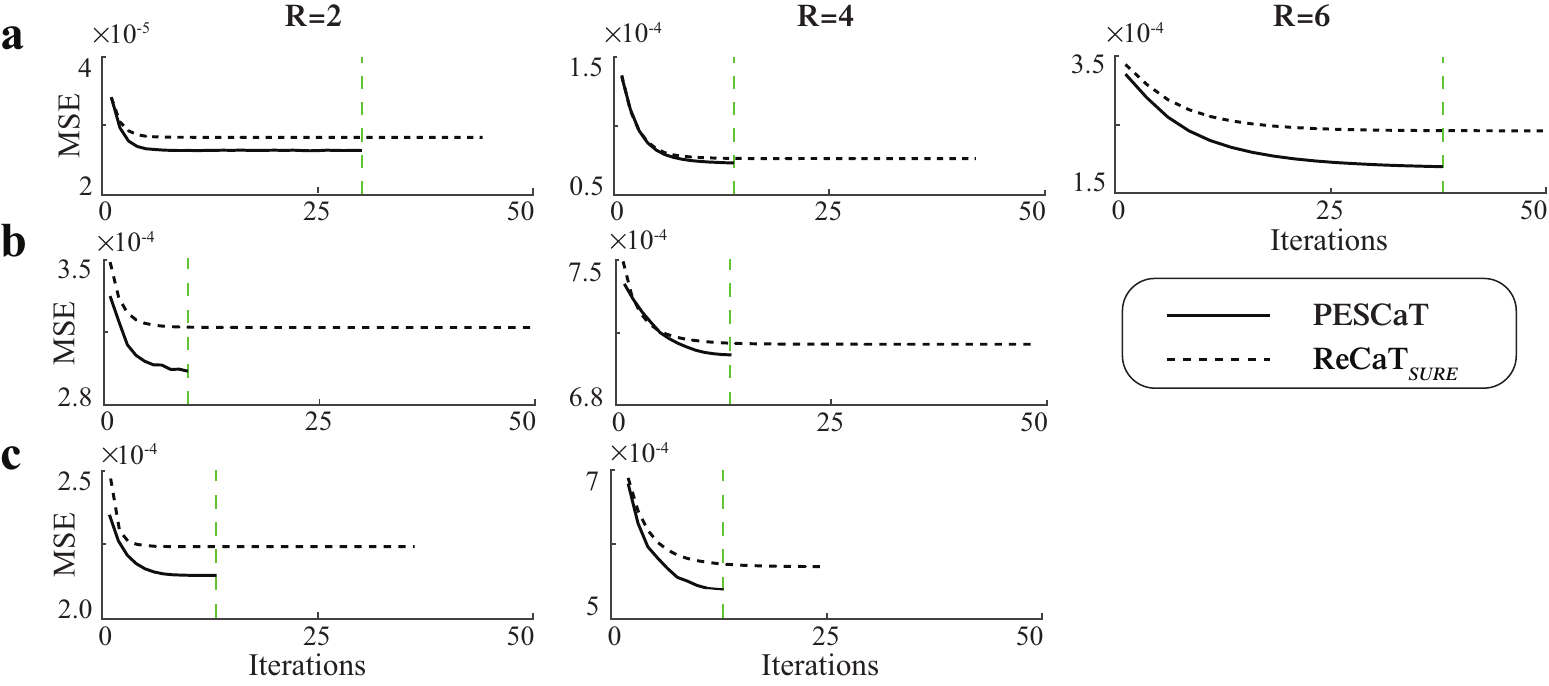}
\setlength{\abovecaptionskip}{-.1cm}
\caption{
Convergence behavior of self-tuning reconstructions was assessed on in vivo \textbf{(a)} bSSFP, \textbf{(b)} T1-weighted, and \textbf{(c)} ToF acquisitions of the brain. 
Mean-squared error (MSE) was calculated between the image reconstructed at each iteration and the fully-sampled reference image. 
The progression of MSE across iterations is shown for a representative cross-section reconstructed using PESCaT \textbf{(solid lines)} and ReCaT\textsubscript{\textit{SURE}} \textbf{(dashed lines)} at R=2 \textbf{(left)}, 4 \textbf{(middle)}, and 6 \textbf{(right)}. 
Reconstructions were stopped once convergence criteria were reached (see \textit{Methods}). 
The iterations at which PESCaT converged are indicated (dashed green lines). 
In all cases, PESCaT converges in a significantly smaller number of iterations, and it converges to a solution with lower MSE than ReCaT\textsubscript{\textit{SURE}}.   
}
\label{fig:convergence}
\end{figure*}

\begin{table}[]
\centering
\label{tab:phantom}
\begin{threeparttable}
\caption{Peak signal-to-noise ratio (PSNR) for simulated phantom}
\footnotesize
\begin{tabular}{llll}
\hline
\multicolumn{1}{c}{} & \multicolumn{1}{c}{\textbf{R=2}}        & \multicolumn{1}{c}{\textbf{R=4}}        & \multicolumn{1}{c}{\textbf{R=6}}      \\ \hline
\textbf{Brute-force}               & \textit{30.29$\pm$0.24} & \textit{28.16$\pm$0.27}  & \textit{27.47$\pm$0.22} \\
\hline
\textbf{PESCaT}                     & \textbf{29.56$\pm$0.34} & \textbf{26.68$\pm$0.24} & \textbf{26.25$\pm$0.19} \\ 
\textbf{ReCaT\textsubscript{\textit{SURE}}}                  & 29.44$\pm$0.21 & 25.52$\pm$0.23 & 24.72$\pm$0.16 \\
\textbf{Early stop}                    & 27.67$\pm$0.26 & 24.96$\pm$0.23 & 23.66$\pm$0.21 \\ 
\hline
\end{tabular}
\begin{tablenotes}\footnotesize
\item PSNR was measured between the reconstructed image and a fully-sampled reference image. 
Measurements were obtained for brute-force, PESCaT, ReCaT\textsubscript{\textit{SURE}} and ReCaT\textsubscript{\textit{SURE}} with early stop methods. 
Results are reported as mean$\pm$std across five cross-sections. 
\end{tablenotes}
\end{threeparttable}
\end{table}

\paragraph{Self-tuning regularized ReCaT (ReCaT\textsubscript{\textit{SURE}})}
Our previously proposed multi-coil multi-acquisition method (ReCaT) did not include any regularization parameters \cite{Biyik_Reconstruction_2017}.
We have implemented a variant of ReCaT incorporating sparsity and TV regularization terms where the regularization parameters are automatically selected using the data.
This reconstruction method iteratively synthesizes unacquired data as weighted combinations of collected data across coils and/or acquisitions. 
It uses sparsity and TV projections to enforce image sparsity.
At each iteration, the regularization parameter for the sparsity term is selected based on the SURE criterion.
The regularization parameter for the TV term is selected based on the local standard deviation of the reconstructed image from the previous iteration.

An alternating projections onto sets algorithm was used in ReCaT\textsubscript{\textit{SURE}} to solve the reconstruction problem cast in (\ref{pescat}). 
ReCaT\textsubscript{\textit{SURE}} used a single sparsity regularization parameter for all subbands and levels of wavelet coefficients.
The sparsity regularization parameter was determined via a line search over the range $[2\times 10^{-5},2\times 10^{-1}]$. 
The TV regularization parameter was taken as one-third of the median local standard deviation \cite{Ilicak_Parameter_2017}. 
All remaining reconstruction parameters were kept identical to PESCaT.

\paragraph{ReCaT\textsubscript{\textit{SURE}} with early stop} The projections performed in each iteration of PESCaT do not involve any line searches, and therefore they are more efficient compared to ReCaT\textsubscript{\textit{SURE}}. 
To enable a fair comparison, we implemented a variant of ReCaT\textsubscript{\textit{SURE}} that was stopped once the total reconstruction time reached that of PESCaT. 
All reconstruction parameters except the total number of iterations were kept identical to ReCaT\textsubscript{\textit{SURE}}.

\paragraph{ReCaT with empirically-tuned parameters (ReCaT\textsubscript{\textit{fixed}})} To demonstrate the effects of prescribing separate regularization parameters for different subbands/levels or coils/acquisitions in PESCaT, we implemented a variant of ReCaT with a single sparsity parameter across all subbands/levels and a single TV parameter across all coils/acquisitions.
Similar to PESCaT, this reconstruction method iteratively synthesizes unacquired data as weighted combinations of collected data across coils and/or acquisitions.
ReCaT\textsubscript{\textit{fixed}} was tuned using held-out data.
The sparsity and TV parameters were independently varied across a broad range $[10^{-5},0.5]$.
Separate reconstructions were obtained for each parameter set, and reconstruction quality was taken as peak signal-to-noise ratio (PSNR) between the reconstructed image and the fully-sampled reference image.
The parameter set that yielded the maximum PSNR was selected. Sparsity and TV parameters were fixed across iterations. All remaining reconstruction parameters were kept identical to PESCaT.

\paragraph{Brute-force reconstruction} To evaluate the success of PESCaT in selecting the optimal parameters, a brute-force reconstruction was implemented to solve the problem in (\ref{pescat}). 
The brute-force method used a constant set of regularization parameters across iterations. 
The sparsity and TV parameters were independently varied across the range $[10^{-5},0.5]$. 
Separate reconstructions were obtained for each parameter set, and reconstruction quality was taken as PSNR between the reconstructed image and the fully-sampled reference image. 
The parameter set that yielded the maximum PSNR was selected.
 All remaining reconstruction parameters were kept identical to PESCaT.

\paragraph{ESPIRiT with PES parameter tuning (PESSPIRiT)} To compare the performance of PESCaT against conventional parallel imaging, we implemented a variant of ESPIRiT \cite{Uecker_ESPIRiT_2014} that included sparsity and TV regularization terms.
Similar to ESPIRiT, this method iteratively reconstructs images based on coil sensitivities estimated from central calibration data.
In each iteration, the sparsity and TV regularization parameters were tuned using PES.
Two other variants, PESSPIRiT with only the sparsity regularization (PESSPIRiT\textsubscript{\textit{$\ell_1$}}) and PESSPIRiT with only the TV regularization (PESSPIRiT\textsubscript{\textit{TV}}) were also implemented.
In all variants, the stopping criterion was identical to PESCaT to enable a fair comparison.

\begin{table*}[]
\centering
\begin{threeparttable}
\caption{PSNR and NRMSE for in vivo bSSFP dataset}
\footnotesize
\begin{tabular}{lllllll}
\hline
& \multicolumn{2}{c}{\textbf{R=2}}             & \multicolumn{2}{c}{\textbf{R=4}}             & \multicolumn{2}{c}{\textbf{R=6}}          \\ 
\hline
         & \multicolumn{1}{c}{\footnotesize{PSNR}} & \multicolumn{1}{c}{ \footnotesize{NRMSE$\times10^{3}$} }& \multicolumn{1}{c}{\footnotesize{PSNR}} & \multicolumn{1}{c}{\footnotesize{NRMSE$\times10^{3}$}} & \multicolumn{1}{c}{\footnotesize{PSNR} }& \multicolumn{1}{c}{\footnotesize{NRMSE$\times10^{3}$} }\\
          \cline{2-7}
\textbf{Brute-force}                            & \textit{44.31$\pm$0.72}          &  \textit{7.31$\pm$0.23}         &\textit{ 40.21$\pm$0.78}          &  \textit{11.18$\pm$0.45}         &  \textit{37.62$\pm$0.61}         &  \textit{16.79$\pm$0.56}         \\
\cline{1-7}
\textbf{PESCaT}                                 & 43.93$\pm$0.65          &  8.16$\pm$0.27         & \textbf{39.64$\pm$0.61} &  \textbf{11.93$\pm$0.43}& \textbf{36.72$\pm$0.29} &  \textbf{18.36$\pm$0.34}         \\
\textbf{ReCaT\textsubscript{\textit{fixed}}}    & \textbf{44.11$\pm$0.62} &  \textbf{7.57$\pm$0.33}& 39.08$\pm$0.47          &  12.59$\pm$0.48         & 36.51$\pm$0.48          &  18.45$\pm$0.95        \\
\textbf{ReCaT\textsubscript{\textit{SURE}}}     & 42.20$\pm$0.78          &  10.09$\pm$0.59         & 37.89$\pm$0.76          &  16.19$\pm$0.79        & 35.15$\pm$0.59          &  20.44$\pm$0.82        \\
\textbf{Early stop}                             & 41.83$\pm$0.66          &  10.37$\pm$0.52         & 36.82$\pm$0.63          &  18.18$\pm$0.77        & 34.01$\pm$0.32          &  22.08$\pm$0.31        \\ 
\textbf{PESSPIRiT}                              & 43.37$\pm$0.44          &  8.80$\pm$0.35         & 38.39$\pm$0.59          &  14.37$\pm$0.84        & 35.06$\pm$0.35          &  20.56$\pm$0.71         \\ 
\textbf{PESSPIRiT\textsubscript{\textit{$\ell_1$}}}& 41.11$\pm$0.62          &  10.84$\pm$0.62         & 35.96$\pm$0.47          &  19.36$\pm$0.86        & 33.15$\pm$0.55          &  23.05$\pm$1.40        \\ 
\textbf{PESSPIRiT\textsubscript{\textit{TV}}}   & 42.41$\pm$0.67          &  9.61$\pm$0.59         & 35.57$\pm$0.68          &  19.88$\pm$1.35        & 32.24$\pm$0.61          &  24.28$\pm$1.69        \\
 \hline
\end{tabular}
\begin{tablenotes}\footnotesize
\item PSNR and NRMSE were measured between the reconstructed image and a fully-sampled reference image. 
Measurements were obtained for brute-force, PESCaT, ReCaT\textsubscript{\textit{SURE}}, ReCaT\textsubscript{\textit{SURE}} with early stop, ReCaT\textsubscript{\textit{fixed}}, and variants of PESSPIRiT methods. 
Results are averaged across three subjects, and reported as mean$\pm$std across five cross-sections.
\end{tablenotes}
\label{tab:vivo_ssfp}
\end{threeparttable}
\end{table*}

 \vspace{-8pt}
\subsection{Simulations}
Simulations were performed using a realistic brain phantom at $0.5$ mm isotropic resolution (\url{http://www.bic.mni.mcgill.ca/brainweb}).
 Phase-cycled bSSFP signals were assumed with T1/T2: 3000/1000 ms for cerebrospinal fluid, 1200/250 ms for blood, 1000/80 ms for white matter, 1300/110 ms for gray matter, 1400/30 ms for muscle, and 370/130 ms for fat \cite{Ilicak_Profile_2017}. 
Single-coil three-dimensional (3D) acquisitions were assumed with TR/TE=5.0/2.5 ms, flip angle=$45^{\circ}$, and phase-cycling increments $\Delta\phi$=$2\pi\frac{[0:1:N-1]}{N}$. 
 We used a simulated field inhomogeneity distribution corresponding to an off-resonance shift with zero mean and 62 Hz standard deviation. 
A bivariate Gaussian noise was added to simulated acquisitions to attain signal-to-noise ratio (SNR)=20, where SNR was taken as the ratio of the mean power in the phantom image to the noise variance.
 Data were undersampled by a factor (R) of 2, 4, and 6 in the two phase-encode directions using disjoint, variable density random undersampling \cite{ukur_Accelerated_2015} and normalized so that zero-filled density compensated k-space data had unity norm \cite{Lustig_Sparse_2007}.
 Reconstruction quality was taken as PSNR between reconstructions and a fully-sampled reference. 
 To prevent bias, the $98^{th}$ percentile of image intensities were adjusted to $[0,1]$. 
 PSNR values were then averaged across five central axial cross-sections.

To examine the effect of noise on optimal regularization parameters, we performed experiments on the simulated brain phantom where the noise level was systematically varied. 
 The simulations output single-coil single-acquisition brain images with SNR varying in the range $[5,25]$. 
 Data were undersampled by R=2, 4, and 6 in the two phase-encode directions using disjoint, variable density random undersampling. Multiple separate reconstructions were obtained for each undersampled dataset via ReCaT\textsubscript{\it{fixed}}, while $\ell_1$ and TV regularization parameters were independently varied in the range $[0.001,0.1]$. 
 At each SNR level, fully-sampled data were used as reference. 
 PSNR was measured between the reconstructions and the reference. 
 The optimal regularization parameters were selected according to PSNR.

\begin{table*}[]
\centering
\begin{threeparttable}
\caption{PSNR and NRMSE for in vivo T1-weighted dataset}
\footnotesize
\begin{tabular}{lllllll}
\hline
  & \multicolumn{2}{c}{\textbf{R=2}}             & \multicolumn{2}{c}{\textbf{R=4}}          \\ 
  \hline
         & \multicolumn{1}{c}{\footnotesize{PSNR} }& \multicolumn{1}{c}{\footnotesize{NRMSE$\times10^{3}$}} & \multicolumn{1}{c}{\footnotesize{PSNR}}  & \multicolumn{1}{c}{\footnotesize{NRMSE$\times10^{3}$}} \\ 
          \cline{2-5}
\textbf{Brute-force}                            & \textit{36.75$\pm$0.55}          &  \textit{18.25$\pm$0.89}         & \textit{32.15$\pm$0.41}          &  \textit{23.75$\pm$0.85}         \\
\cline{1-5}
\textbf{PESCaT}                                 & \textbf{35.62$\pm$0.95} &  \textbf{19.86$\pm$0.99}& \textbf{31.44$\pm$1.09} &  \textbf{27.75$\pm$1.61}\\
\textbf{ReCaT\textsubscript{\textit{fixed}}}    & 35.27$\pm$0.67          &  20.02$\pm$0.85         & 30.89$\pm$0.63          &  29.59$\pm$1.31        \\
\textbf{ReCaT\textsubscript{\textit{SURE}}}     & 35.02$\pm$0.93          &  20.64$\pm$0.94         & 30.67$\pm$0.93          &  30.10$\pm$1.57        \\
\textbf{Early stop}                             & 34.64$\pm$1.03          &  21.47$\pm$0.95         & 30.03$\pm$1.06          &  30.81$\pm$1.65        \\ 
\textbf{PESSPIRiT}                              & 35.21$\pm$0.74          &  20.14$\pm$1.10         & 29.92$\pm$0.63          &  31.66$\pm$1.60        \\ 
\textbf{PESSPIRiT\textsubscript{\textit{$\ell_1$}}}& 31.75$\pm$0.66          &  27.37$\pm$1.42         & 27.39$\pm$0.80          &  34.37$\pm$2.72        \\ 
\textbf{PESSPIRiT\textsubscript{\textit{TV}}}   & 34.93$\pm$0.69          &  21.04$\pm$1.03         & 29.48$\pm$0.60          &  31.83$\pm$1.61        \\ 
\hline
\end{tabular}
\begin{tablenotes}\footnotesize
\item PSNR and NRMSE were measured between the reconstructed image and a fully-sampled reference image. 
Measurements were obtained for brute-force, PESCaT, ReCaT\textsubscript{\textit{SURE}}, ReCaT\textsubscript{\textit{SURE}} with early stop, ReCaT\textsubscript{\textit{fixed}}, and variants of PESSPIRiT methods. 
Results are averaged across three subjects, and reported as mean$\pm$std across five cross-sections.
\end{tablenotes}
\label{tab:vivo_t1}
\end{threeparttable}
\end{table*}

To examine the reliability of the epigraph scaling parameters against noise, reconstructions of the brain phantom were obtained at three separate levels of SNR $=10,18, 25$. Meanwhile, $\beta_{\ell_1}$ was varied in the range $[0.05,0.6]$ and $\beta_{TV}$ was varied in the range $[0.1,1]$. To examine the reliability of the epigraph scaling parameters against variations in the level of detail and spatial resolution, we performed experiments on a simulated numerical phantom. A circular phantom of radius 125 voxels (for a 256$\times$256 field of view) was designed with the background resembling muscle tissue and vertical bright bars of width 12 and height [190, 220, 238, 238, 220, 190] voxels resembling blood vessels (Supp. Fig. 1)\footnote{supplementary materials are available in the supplementary files /multimedia tab.}. Phase-cycled bSSFP signals were assumed with T1/T2: 870/47 ms for muscle, and 1273/259 ms for blood. Three dimensional acquisitions were assumed with TR/TE=4.6/2.3 ms, flip angle=$60^{\circ}$, and phase-cycling increments $\Delta\phi$=$2\pi\frac{[0:1:N-1]}{N}$. A simulated field inhomogeneity distribution corresponding to an off-resonance shift with zero mean and 62 Hz standard deviation was used. Level of detail was varied from low to high by incrementally placing [1, 3, 6] vertical bars in the phantom. Spatial resolution was varied from low to high by low-pass filtering k-space data to select circular regions of radius [20, 55, 125] voxels. Reconstructions were obtained while $\beta_{\ell_1}$ and $\beta_{TV}$ were varied in the range $[0.05,0.5]$. 

 \vspace{-12pt}
\subsection{In vivo experiments}
Experiments were performed to acquire 3D multi-coil multi-acquisition phase-cycled bSSFP, and multi-coil single-acquisition T1-weighted and time-of-flight (ToF) angiography data in the brain. 
Data were collected on a 3T Siemens Magnetom scanner (maximum gradient strength of 45 mT/m and slew rate of 200 T/m/s).
bSSFP and ToF data were collected using a 12-channel receive-only head coil that was hardware compressed to 4 channels.
 T1-weighted data were collected using a 12-channel receive-only head coil.  
 Separate bSSFP datasets were also collected using a 32-channel head coil.
  Balanced SSFP data were acquired using a bSSFP sequence with the following parameters: flip angle=$30^{\circ}$, TR/TE=8.08/4.04 ms, field-of-view (FOV)=218 mm $\times$ 218 mm, matrix size of 256 $\times$ 256 $\times$ 96, resolution of 0.9 mm $\times$ 0.9 mm $\times$ 0.8 mm, right/left readout direction, and N=8 separate acquisitions with phase-cycling values in the range $[0,2\pi)$ in equispaced intervals. 
Total acquisition time for the bSSFP sequence was 20:56.
T1-weighted data were acquired using an MP-RAGE sequence with the parameters: flip angle=$9^{\circ}$, TR/TE=2300/2.98 ms, TI=900 ms, FOV= 256 mm $\times$ 240 mm, matrix size of 256 $\times$ 240 $\times$ 160, resolution of  1.0 mm $\times$ 1.0 mm $\times$ 1.2 mm, and superior/inferior readout direction.
Total acquisition time for the MP-RAGE sequence was 9:14.
ToF angiograms were acquired using a multiple overlapping thin-slab acquisition (MOTSA) sequence with parameters:  flip angle=$18^{\circ}$, TR/TE=38/3.19 ms, FOV=$204$ mm $\times$ 204 mm, matrix size of 256 $\times$ 256 $\times$ 75, isotropic resolution of $0.8$ mm, and anterior/posterior readout direction.
Total acquisition time for the MOTSA sequence was 14:16.
The imaging protocols were approved by the local ethics committee, and all six participants gave written informed consent.

Phase-cycled bSSFP acquisitions with 4 channels were retrospectively undersampled at R=2, 4, and 6. Following phase-cycles were selected: $\Delta\phi = 2\pi\frac{[0:1:N-1]}{N}$ for N=2 and 4, and $[0,\frac{\pi}{2},\frac{3\pi}{4},\pi,\frac{5\pi}{4},\frac{7\pi}{4}]$ for N=6. 
For this bSSFP dataset, N=R was used. 
T1-weighted and ToF acquisitions were retrospectively undersampled at R=2 and 4 (note that in these cases N=1). Undersampling was performed across the two phase encode directions: superior/inferior and anterior/posterior for bSSFP, right/left and anterior/posterior for T1-weighted, superior/inferior and right/left for ToF. Data were normalized so that zero-filled density compensated k-space data had unity norm. 
 Entire volumes were reconstructed, five axial cross-sections equispaced across the entire brain were selected for quantitative assessment.
PSNR and normalized root mean-squared error (NRMSE) measurements were averaged across cross-sections.

To investigate the convergence behavior of PESCaT, we studied the evolution of the three cost terms in (\ref{pescat}) separately (Supp. Fig. 2).
Normalized cost terms associated with calibration consistency, sparsity, and TV terms at the end of each iteration were plotted across iterations.
In all datasets, all cost terms diminish smoothly.

To optimize epigraph scaling constants for $\ell_1$- and TV-norm functions, PESCaT was performed on data acquired from three subjects reserved for this purpose. 
Volumetric reconstructions were performed at R=2, 4, and 6 for bSSFP datasets, and R=2 and 4 for T1-weighted and ToF datasets. 
Separate reconstructions were obtained while $\beta_{\ell_1}$ was varied in the range $[0.1,1]$, and $\beta_{TV}$ was varied in the range $[0.05,0.6]$. 
PSNR was measured between the reconstructed and fully-sampled reference images (Supp. Figs. 3, 4).
Consistently across subjects and different types of datasets, PSNR values within 95\% of the optimum value were maintained in the range $\beta_{\ell_1} = [0.1,0.3]$, and $\beta_{TV}= [0.2,0.4]$. 
Near-optimal PSNR values were attained around $\beta_{\ell_1} = 0.2$ and $\beta_{TV}=0.3$. 
Thus, these scaling constants were prescribed for reconstructions thereafter.

To demonstrate the reconstruction performance of PESCaT at high acceleration rates, phase-cycled bSSFP acquisitions with 32 channels were analyzed. This bSSFP dataset was retrospectively undersampled at R= 8, 10 (where N=8). Entire volumes were reconstructed, and PSNR and NRMSE measurements were averaged across five axial cross-sections. 

\begin{table*}[]
\centering
\begin{threeparttable}
\caption{PSNR and NRMSE for in vivo ToF dataset}
\footnotesize
\begin{tabular}{lllllll}
\hline
 & \multicolumn{2}{c}{\textbf{R=2}}             & \multicolumn{2}{c}{\textbf{R=4}}          \\ 
 \hline
         & \multicolumn{1}{c}{\footnotesize{PSNR}} & \multicolumn{1}{c}{\footnotesize{NRMSE$\times10^{3}$}} & \multicolumn{1}{c}{\footnotesize{PSNR}} & \multicolumn{1}{c}{\footnotesize{NRMSE$\times10^{3}$}} \\ 
         \cline{2-5}
\textbf{Brute-force}                            & \textit{36.57$\pm$1.61}          &  \textit{18.61$\pm$0.71}         & \textit{33.08$\pm$1.55 }         &  \textit{22.75$\pm$0.97}        \\
\cline{1-5}
\textbf{PESCaT}                                 & \textbf{36.32$\pm$1.14} &  \textbf{18.79$\pm$0.57}& \textbf{31.86$\pm$1.21} &  \textbf{27.55$\pm$1.03}\\
\textbf{ReCaT\textsubscript{\textit{fixed}}}    & 35.86$\pm$0.60          &  19.41$\pm$0.86         & 30.51$\pm$0.60          &  30.17$\pm$1.60        \\
\textbf{ReCaT\textsubscript{\textit{SURE}}}     & 35.55$\pm$1.25          &  19.97$\pm$0.94         & 31.34$\pm$1.13          &  28.40$\pm$1.57        \\
\textbf{Early stop}                             & 35.46$\pm$1.12          &  20.14$\pm$0.95         & 27.69$\pm$1.19          &  28.71$\pm$1.65        \\ 
\textbf{PESSPIRiT}                              & 35.86$\pm$0.57          &  19.42$\pm$0.90         & 30.83$\pm$0.47          &  29.72$\pm$1.41        \\ 
\textbf{PESSPIRiT\textsubscript{\textit{$\ell_1$}}}& 32.00$\pm$0.55          &  24.89$\pm$1.46         & 27.22$\pm$0.58          &  37.84$\pm$2.71        \\ 
\textbf{PESSPIRiT\textsubscript{\textit{TV}}}   & 35.66$\pm$0.53          &  19.72$\pm$0.84         & 30.41$\pm$0.64          &  30.42$\pm$1.89        \\ 
\hline
\end{tabular}
\begin{tablenotes}\footnotesize
\item PSNR and NRMSE were measured between the reconstructed image and a fully-sampled reference image. 
Measurements were obtained for brute-force, PESCaT, ReCaT\textsubscript{\textit{SURE}}, ReCaT\textsubscript{\textit{SURE}} with early stop, ReCaT\textsubscript{\textit{fixed}}, and variants of PESSPIRiT methods. 
Results are averaged across three subjects, and reported as mean$\pm$std across five cross-sections.
\end{tablenotes}
\label{tab:vivo_tof}
\end{threeparttable}
\end{table*}
 \vspace{-8pt}
\section{Results}
 \vspace{-2pt}
\subsection{Simulations}

MRI data may show differential noise and structural characteristics for separate coils and acquisitions, or for separate wavelet subbands and levels. 
 In turn, the optimal regularization parameters can vary across each of these dimensions. 
 To test this prediction, we performed experiments on the simulated brain phantom, where the noise level was systematically varied and ReCaT\textsubscript{\it{fixed}} reconstructions were performed. For both $\ell_1$ and TV regularization, the optimal regularization parameters show a clear increasing trend as SNR is lowered (Supp. Fig. 5). These results suggest that prescribing a fixed parameter can cause performance loss when a good compromise cannot be achieved across subbands/levels or coils/acquisitions. It can also render the reconstruction more susceptible to deviations from the optimal value of the regularization parameter. 

In contrast, PESCaT uses only two global parameters to control the overall sparsity of the solutions in wavelet domain ($\beta_{l_1}$) and TV domain ($\beta_{TV}$). 
 Given these scaling parameters, regularization parameters for individual subbands/levels and coils/acquisitions are determined adaptively in a data-driven manner. To examine the reliability of the scaling parameters against noise, reconstructions were obtained at varying SNR levels. The PSNR curves as a function of $\beta_{\ell_1}$ and $\beta_{TV}$ demonstrate substantial flatness, yielding near-optimal performance across the entire range of values examined (Supp. Fig. 6). To further examine the reliability of the scaling parameters against variations in the level of detail and spatial resolution, reconstructions were obtained at low, medium and high levels of detail and resolution. Supp. Fig. 7 displays PSNR across $\beta_{\ell_1}$ and Supp. Fig. 8 displays PSNR across $\beta_{TV}$ values. Again, PSNR curves as a function of $\beta_{\ell_1}$ and $\beta_{TV}$ demonstrate substantial flatness, yielding near-optimal performance across the entire range of values examined.

Following these basic demonstrations, PESCaT was performed on bSSFP acquisitions of a simulated brain phantom. Representative reconstructions and error maps for PESCaT and ReCaT\textsubscript{\textit{SURE}} with R=6 are shown in Fig. \ref{fig:phantom}. 
PESCaT yields reduced error across the FOV compared to ReCaT\textsubscript{\textit{SURE}}. 
This improvement with PESCaT becomes further noticeable when ReCaT\textsubscript{\textit{SURE}} is stopped early to match its reconstruction time to PESCaT. 
Quantitative assessments of image quality at R=2, 4, and 6 are listed in Table I.  
Among all techniques tested, PESCaT achieves the most similar performance to the time-consuming brute-force reconstruction. 
On average, PESCaT improves PSNR by $0.87\pm0.74$ dB over ReCaT\textsubscript{\textit{SURE}} and by $1.87\pm0.73$ dB over ReCaT\textsubscript{\textit{SURE}} with early stop (mean$\pm$std. across five cross-sections, average of R=2, 4, 6).
Note that the proposed method attains near-optimal performance while enabling improved computational efficiency. 
The average reconstruction time per slice is $27\pm5$ s for ReCaT\textsubscript{\textit{SURE}} and only $7\pm4$ s for PESCaT, resulting in a 4-fold gain in efficiency for the phantom dataset.

 \vspace{-8pt}
 \subsection{In vivo experiments}
We first examined the evolution of the cost terms during PESCaT reconstruction of in vivo bSSFP and T1-weighted datasets  (Supp. Figs. 9, 10).
Both $\ell_1$ and TV cost terms decrease towards later iterations indicating that the images better conform to a compressible representation.

Next, PESCaT was demonstrated for in vivo bSSFP, T1-weighted, and ToF imaging of the brain. 
Representative reconstructions with R=6 for bSSFP and R=4 for T1-weighted and ToF acquisitions are displayed in Figs. \ref{fig:ssfp}, \ref{fig:t1}, and \ref{fig:tof}. 
Representative reconstructions of individual phase cycles in the bSSFP dataset, and of cross-sections in the ToF dataset are shown in Supp. Fig. 11.
Overall, PESCaT and ReCaT\textsubscript{\textit{SURE}} reconstructions perform similar to the brute-force optimized reconstructions. 
Yet, PESCaT yields slightly lower levels of residual aliasing in comparison to ReCaT\textsubscript{\textit{SURE}}, and this difference is particularly noticeable for visualization of small vessels in ToF images (Fig. \ref{fig:tof}). 
The improvement in reconstruction quality with PESCaT is more prominent when ReCaT\textsubscript{\textit{SURE}} is stopped early to match its reconstruction time to PESCaT. 

Quantitative assessments of the in vivo reconstructions are listed in  Tables \ref{tab:vivo_ssfp}, \ref{tab:vivo_t1}, and \ref{tab:vivo_tof}. 
For all datasets and R, PESCaT yields the closest performance to the brute-force reconstruction among alternative self-tuning methods. 
For bSSFP datasets, PESCaT improves PSNR by $1.23\pm0.29$ dB over ReCaT\textsubscript{\textit{SURE}} and by $2.55\pm0.51$ dB over ReCaT\textsubscript{\textit{SURE}} with early stop (mean$\pm$std. across three subjects, average of R=2, 4, 6). 
For T1-weighted datasets, PESCaT improves PSNR by $0.71\pm 0.25$ dB over ReCaT\textsubscript{\textit{SURE}} and by $1.21\pm0.43$ dB over ReCaT\textsubscript{\textit{SURE}} with early stop (mean$\pm$std. across three subjects, average of R=2, 4). 
For ToF datasets, PESCaT improves PSNR by $0.72\pm0.46$ dB over ReCaT\textsubscript{\textit{SURE}} and by $0.94\pm0.51$ dB over ReCaT\textsubscript{\textit{SURE}} with early stop (mean$\pm$std. across three subjects, average of R=2, 4). Compared to empirically-tuned ReCaT\textsubscript{\textit{fixed}}, PESCaT improves PSNR  by $0.20\pm0.37$ dB for bSSFP datasets, by $0.45\pm0.14$ dB for T1-weighted datasets, and by $0.91\pm0.63$ dB for ToF datasets (mean$\pm$std. across three subjects, average of R=2, 4, 6 for bSSFP, average of R=2, 4 for T1-weighted and ToF datasets. 
Because both methods were allowed to optimize parameters in training subjects, these results suggest that selecting different regularization parameters for each coil/acquisition/subband/level improves reconstruction performance. Compared to PESSPIRiT, PESCaT improves PSNR  by $1.16\pm0.55$ dB for bSSFP datasets, by $0.97\pm0.78$ dB for T1-weighted datasets, and by $0.76\pm0.40$ dB for ToF datasets (mean$\pm$std. across three subjects, average of R=2, 4, 6 for bSSFP, average of R=2, 4 for T1-weighted and ToF datasets). Performance enhancement is even more prominent compared to PESSPIRiT variants that only include sparsity or TV regularization. 

To assess the computational efficiency of self-tuning methods, representative reconstructions were performed for a single cross-section of in vivo bSSFP, T1-weighted, and ToF acquisitions. 
The true MSE between the reconstructed and fully-sampled reference images were recorded across iterations of PESCaT and ReCaT\textsubscript{\textit{SURE}}. 
MSE curves across iterations are displayed in Fig. \ref{fig:convergence}. 
Compared to ReCaT\textsubscript{\textit{SURE}}, the proposed method converges to a lower MSE value for all R and datasets. 
Furthermore, PESCaT reduces the number of iterations by 43.3\% for bSSFP (average over R=2, 4, 6), 74.5\% for T1-weighted (average over R=2, 4) and 53.2\% for ToF (average over R=2, 4) datasets. 
Note that each iteration of PESCaT performs more efficient geometric projections without explicit parameter searches. The reconstruction times for PESCaT and alternative methods are listed in Supp. Table I. On average, the reconstruction time of ReCaT\textsubscript{\textit{SURE}} was $1641\pm45$ s for bSSFP, $1799\pm66$ s for T1-weighted, and $565\pm58$ s for ToF datasets (mean$\pm$std. across five cross-sections, average over R=2, 4 for T1-weighted and ToF imaging; R=2, 4, 6 for bSSFP imaging). In contrast, the reconstruction time of PESCaT was merely $164\pm25$ s for bSSFP, $196\pm44$ s for T1-weighted, and $159\pm32$ s for ToF datasets. These results suggest that PESCaT offers up to 10-fold gain in efficiency compared to the alternative self-tuning method ReCaT\textsubscript{\textit{SURE}}. While PESSPIRiT  yields similar reconstruction times and ReCaT\textsubscript{\textit{fixed}} slightly reduces reconstruction times compared to PESCaT, both methods yield inferior reconstruction quality. 

Lastly, reconstruction performance of PESCaT was demonstrated at higher acceleration rates using the 32-channel bSSFP datasets (Supp. Fig. 12 and Supp. Table II).
The proposed method improves PSNR by $0.14\pm0.04$ compared to ReCaT\textsubscript{\textit{fixed}}, by $1.59\pm0.45$ compared to ReCaT\textsubscript{\textit{SURE}}, by $3.77\pm0.61$ compared to ReCaT\textsubscript{\textit{SURE}} with early stop, and by $4.08\pm0.55$ over PESSPIRiT (mean$\pm$std. across three subjects, average of R=8, 10).
These results help demonstrate the utility of PESCaT in enabling higher acceleration factors when using modern coil arrays.
 \vspace{-8pt}
\section{Discussion}
In this study, we have proposed a new self-tuning method for CS reconstruction of single-coil multi-acquisition, multi-coil single-acquisition, and multi-coil multi-acquisition datasets. 
The proposed method performs sparsity projections across coils and acquisitions to penalize the $\ell_1$-norm of wavelet coefficients, and TV projections to penalize the finite-differences gradients of image coefficients. 
 Separate sparsity regularization parameters are selected at each wavelet subband and level, and separate TV regularization parameters are selected at each coil and acquisition. 
 Efficient projections onto the boundary of the epigraph sets of the $\ell_1$-norm and TV-norm functions are used to simultaneously calculate the projections themselves and automatically determine the relevant regularization parameters. PESCaT does not have any constraints regarding the number of acquisitions or coils. Therefore, it can be readily applied to both single-acquisition and multi-acquisition datasets regardless of the number of coils available. PESCaT also offers flexibility regarding the inclusion of regularization terms. Because the algorithm has a modular structure with respect to individual calibration, sparsity, and TV projections, it is possible to omit either TV or sparsity regularization. The proposed method will still work towards a solution at the intersection of the remaining projection sets.

In a recent study, we proposed a reconstruction for multi-coil multi-acquisition bSSFP imaging, named ReCaT \cite{Ilicak_Parameter_2017}.
Here, we have implemented a self-tuning version of ReCaT (ReCaT\textsubscript{\textit{SURE}}).
Similar to PESCaT, ReCaT\textsubscript{\textit{SURE}} uses sparsity projections implemented via soft-thresholding and TV projections implemented via iterative clipping. 
However, in ReCaT\textsubscript{\textit{SURE}}, the sparsity regularization parameter was selected via a SURE-based method to minimize the expected reconstruction error.
TV regularization parameter was selected in a data-driven manner based on the local standard deviations within the reconstructed image. 
Since parameter selection in ReCaT\textsubscript{\textit{SURE}} involves line searches over a relevant range of parameters, it can be computationally expensive.  
In contrast, PESCaT leverages highly efficient geometric projections onto epigraph sets to simultaneously select the optimal parameters and calculate the projections. 
Hence, PESCaT enables significant savings in reconstruction time compared to self-tuning methods based on line searches. 
Meanwhile, the main advantage of PESCaT over an empirically-tuned reconstruction that optimizes regularization parameters on training data is that it allows for independent selection of regularization parameters for each coil/acquisition/subband/level. 
The superior reconstruction quality of PESCaT compared to ReCaT\textsubscript{\textit{fixed}} confirms this prediction quantitatively. 

The proposed method includes two epigraph scaling constants $\beta_{\ell_1}$ and $\beta_{TV}$ as free parameters. 
Here we have empirically demonstrated that the optimal scaling constants are highly consistent across individual subjects, across different noise levels and across multiple imaging contrasts of the same anatomy.  
These observations are also complemented by prior work that suggests that the solutions of epigraph sets projections are robust against deviations from optimal scaling constants  \cite{Tofighi_Signal_2014, chierchia2015epigraphical}. 
 It remains to be demonstrated whether the scaling constants are also similar across different anatomies. Still, we expect that PESCaT shows improved robustness against variability in datasets compared to the empirically-tuned ReCaT\textsubscript{\textit{fixed}}. The optimal regularization parameters for ReCaT\textsubscript{\textit{fixed}} showed relatively high variability across the datasets examined in this study (not shown). Thus, ReCaT\textsubscript{\textit{fixed}} might require more careful tuning of regularization parameters, resulting in relatively higher computational overhead.

Further performance improvements might be attained by addressing some limitations of the proposed method. 
For multi-acquisition datasets, significant motion among acquisitions can reduce reconstruction quality. 
A motion-correction projection can be incorporated into the PESCaT algorithm to mitigate artifacts due to the residual motion. 
Second, the proposed method uses a fully-sampled central region in k-space to estimate the tensor interpolation kernel. 
In applications where the acquisition of calibration data is impractical such as spectroscopic and dynamic imaging, calibrationless approaches could be incorporated for improved performance \cite{Trzasko_Calibrationless_2011,Shin_Calibrationless_2014}.
Third, although the epigraph scaling constants $\beta_{\ell_1}$ and $\beta_{TV}$ were optimized over a held-out dataset, it might be possible to automatically select them using parameter selection via SURE or GCV.
This remains an important future research direction toward fully-automated reconstructions.

Here, the alternating projections onto sets algorithm was used to find a solution at the intersection point of the sets corresponding to calibration, sparsity, and TV projections. 
Rapid convergence was observed in all examined cases. 
However, in situations where the intersection between these sets is sparsely populated, more sophisticated algorithms such as alternating direction method of multipliers (ADMM) could be used for fast and effective optimization \cite{boyd2011distributed}.
PESCaT employs projections onto epigraph sets to concurrently select regularization parameters and perform projections. 
As such, it is non-trivial to efficiently adapt the proposed parameter selection to an ADMM-based reconstruction. It remains an important future work to benchmark PESCaT against ADMM coupled with an appropriate parameter-selection strategy.

Projections onto epigraph sets were used to penalize $\ell_1$-norm and TV-norm functions in this study. 
Note that the projection onto convex sets formulation allows penalization of any convex function. 
Thus, the proposed technique could be generalized to include alternative regularizes such as filtered variation or total generalized variation \cite{Kose_Filtered_2012,Knoll_Second_2011}. 
These modifications might allow performance enhancements in applications where standard TV regularization yields undesirable block artifacts. 

In conclusion, PESCaT enables near-optimal image quality while automatically selecting regularization parameters in reconstructions of undersampled MRI datasets. 
Parameter selection for $\ell_1$-norm and TV-norm regularizers and projections onto the $\ell_1$ and TV-balls are performed simultaneously. PESCaT was demonstrated to outperform alternative self-tuning approaches based on SURE in bSSFP, T1-weighted and time-of-flight angiographic imaging. The results presented here demonstrate that PESCaT is a promising method for CS-MRI in routine practice.

 \vspace{-15pt}
\bibliographystyle{IEEEtran}
\bibliography{bibliography}


\end{document}